\documentclass[aps,twocolumn,pra]{revtex4}
\usepackage{epsfig}
\usepackage{graphicx}
\usepackage{amssymb,amsmath}
\usepackage{bbm}
\usepackage{moreverb}
\usepackage[dvips]{color}

\newcommand {\be}{\begin{equation}}
\newcommand {\ee}{\end{equation}}
\newcommand {\ba}{\begin{eqnarray}}
\newcommand {\ea}{\end{eqnarray}}

\begin{document}

\title{Mesoscopic Effects in Quantum Phases of Ultracold Quantum Gases in Optical Lattices}
\author{L. D. Carr$^{1,2}$, M. L. Wall$^1$, D. G. Schirmer$^{1,2}$, R. C. Brown$^{2}$, J. E. Williams$^{2}$, and Charles W. Clark$^{2}$}
\affiliation{$^1$Department of Physics, Colorado School of Mines, Golden, CO 80401, USA}
\affiliation{$^2$Joint Quantum Institute, National Institute of Standards and Technology, Gaithersburg, MD 20899, USA }

\begin{abstract}
We present a wide array of quantum measures on numerical solutions of 1D Bose- and Fermi-Hubbard Hamiltonians for finite-size systems with open boundary conditions.  Finite size effects are highly relevant to ultracold quantum gases in optical lattices, where an external trap creates smaller effective regions in the form of the celebrated ``wedding cake'' structure and the local density approximation is often not applicable.  Specifically, for the Bose-Hubbard Hamiltonian we calculate number, quantum depletion, local von-Neumann entropy, generalized entanglement or Q-measure, fidelity, and fidelity susceptibility; for the Fermi-Hubbard Hamiltonian we also calculate the pairing correlations, magnetization, charge-density correlations, and antiferromagnetic structure factor.  Our numerical method is imaginary time propagation via time-evolving block decimation.  As part of our study we provide a careful comparison of canonical vs. grand canonical ensembles and Gutzwiller vs. entangled simulations.  The most striking effect of finite size occurs for bosons: we observe a strong blurring of the tips of the Mott lobes accompanied by higher depletion, and show how the location of the first Mott lobe tip approaches the thermodynamic value as a function of system size.
\end{abstract}

\maketitle

\section{Introduction}
\label{sec:introduction}
Quantum phase transitions (QPTs) are often treated in the thermodynamic limit, where the concepts of a critical point, critical exponents, and scaling relations are very clear~\cite{sachdev1999}.  However, many physical many-body systems for practical applications these days are finite.  For example, nano-devices can be smaller than 100 nm.  Then a crystal with a lattice constant of a fraction of a nanometer has just 2 or 3 orders of magnitude to work with in scale.  In nuclear physics one has QPTs in finite nuclei which are very far from being in the thermodynamic limit~\cite{caprio2008}.  Experiments on trapped ions, recently used to simulate many body Hamiltonians, normally work with small systems~\cite{kimK2009}.  Ultracold atoms in optical lattices provide a fourth example, where the number of lattice sites is typically 50 to 100 in each direction, so that in one dimension (1D) they are highly mesoscopic.  These systems have been proposed as quantum simulators to solve hard quantum many body problems, including the positive-$U$ Fermi-Hubbard Hamiltonian for general filling~\cite{hofstetter2002,jaksch2004,lewensteinM2007,bloch2008}.  Progress towards this goal requires understanding of how finite-size effects change the quantum phase diagram.  In this paper, we explore such mesoscopic or finite size effects for both Bose- and Fermi-Hubbard Hamiltonians.

%%%%%%%%%%% figure 1 %%%%%%%%%%%
%
\begin{figure}[t]
\begin{center}
\epsfxsize=8cm \epsfysize=6.5 cm \leavevmode \epsfbox{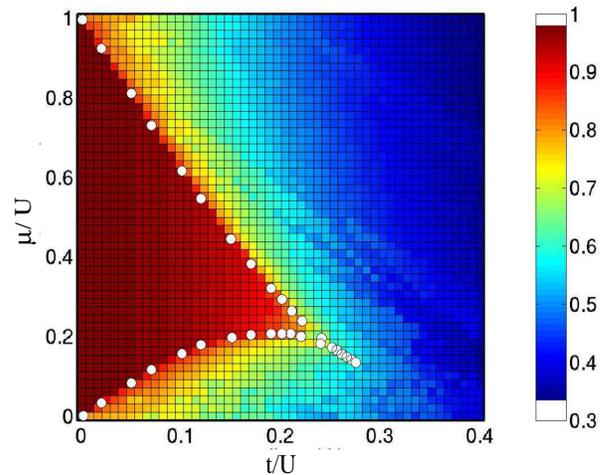}
\caption{\label{fig:williams} (color online) \emph{The first Mott lobe of the Bose-Hubbard Hamiltonian}.  Quantum depletion for a system of interacting bosons on 51 lattice sites at $T=0$ as a function of chemical potential $\mu$ and hopping parameter $t$, both normalized to interaction strength $U$, obtained with Time-Evolving Block Decimation (TEBD) in imaginary time propagation, as discussed below.  The white circular points represent best-converged results of Density Matrix Renormalization Group methods for the boundary of the Mott lobe in the thermodynamic limit~\cite{kuhner2000}.}
\end{center}
\end{figure}

In addition to finite-size effects, ultracold atomic quantum simulators are non-uniform, as ordinarily there is a weak overall harmonic trap keeping atoms from spilling off the edge of the lattice.  This non-uniformity gives rise to a ``wedding-cake'' or tiered structure of Mott plateaus separated by superfluid layers for bosons~\cite{batrouniGG2002,gemelke2009}, and similar layered structure for fermions~\cite{schneiderU2008,jordens2008}.  The non-uniformity creates even smaller regions of Mott insulator or superfluid, often just 10 or 20 sites in length.  For these small regions the quantum phase diagram is substantially different than in the thermodynamic limit: for example, the quantum depletion, (defined precisely below in Sec.~\ref{sec:equations}) shows that the Mott lobe is not closed but instead goes through a very broad crossover into the superfluid regime, as we show in Sec.~\ref{ssec:bose}. Figure~\ref{fig:williams} shows the sharp phase boundary of a Bose Hubbard system in the thermodynamic limit (white circles~\cite{kuhner2000}) and the continuously variable quantum depletion (defined in Sec.~\ref{sec:equations}) obtained in our calculations for 51 sites.  As can be seen in the figure, the tip of the Mott lobe is significantly blurred even for this relatively large lattice.

Approximating the effects of a slowly varying trap by using a local chemical potential in the local density approximation, $\mu=\mu(\vec{r})$, is not very useful for cold atoms in 1D.  The breakdown of the local density approximation has been explored with quantum Monte Carlo and exact diagonalization elsewhere (see~\cite{rigol2009} and references therein); our work, which isolates finite size effects in small uniform regions, is complementary to previous studies which have explored the harmonic trap structure as a whole~\cite{batrouniGG2002,rigol2004,rigol2009}.  The success of quantum simulators relies in part on understanding all possible sources of error~\cite{trotzky2009}, in the same sense as one might do so for a high precision measurement of time-reversal symmetry, for example.  Therefore going beyond the local density approximation~\cite{rigol2009} is key to the eventual success of quantum simulators, especially in lower dimensions.  Our use of TEBD offers a wide array of quantum measures, many of which are not directly accessible to quantum Monte Carlo~\cite{kuhner1998,sengupta2005,fubini2006} or exact 1D methods~\cite{Lieb1968,gu2004}, including various entanglement entropies and fidelity.  Multiple measures provide maximal information, useful for a thorough analysis; in particular, fidelity and fidelity susceptibility make no assumptions about states or symmetries, but simply ask about many body wavefunction overlap in a model-independent way~\cite{venuti2008}.

It has already been shown that the mean-field Gutzwiller approximation, which assumes a product state between sites of the lattice, is not very useful in 1D for Hubbard Hamiltonians~\cite{lewensteinM2007}.  But how far does one have to move away from the Gutzwiller approximation in order to get correct results?  A high level of effort has been put into alternative methods, such as generalized dynamical mean field theory in two dimensions for Bose-Fermi mixtures~\cite{titvinidze2009}.  The convergence parameter in TEBD, which can be couched in terms of the degree of spatial entanglement, allows one to go beyond the Gutzwiller approximation in a methodical manner, and we present a thorough study of this issue.  A second question concerning numerical methods is the use of the canonical ensemble vs. the grand canonical ensemble.  The two only formally give the same result in the thermodynamic limit.  The canonical ensemble is numerically preferred for TEBD and related methods~\cite{daley2004} because the number of Fock states is much smaller and therefore calculations are more efficient.  We explicitly explore the effects of ensemble dependence Hubbard Hamiltonian phase diagrams.

This article is outlined as follows.  In Sec.~\ref{sec:equations} we present the Bose- and Fermi-Hubbard Hamiltonians, the many quantum measures we use, and a brief description of and motivation for our implementation of TEBD.  In Sec.~\ref{sec:bosefermi} we present and discuss our numerical results for mesoscopic effects in ground states of the Bose- and Fermi-Hubbard Hamiltonians, respectively.  In Sec.~\ref{sec:compare} we compare canonical to grand canonical ensembles and Gutzwiller to entangled TEBD simulations; these comparisons also serve as a convergence study.  Finally, in Sec.~\ref{sec:conclusion}, we conclude.

\section{Models, Measures, and Methods}
\label{sec:equations}

We treat two fundamental Hamiltonians for ultracold quantum gases in optical lattices, the Bose- and Fermi-Hubbard Hamiltonians.  In both cases we take the tight-binding lowest-band approximation.  The Bose-Hubbard Hamiltonian is
\ba
\hat{H_b}&=& -t\sum_{\langle i,j\rangle}(\hat{b}^{\dagger}_i \hat{b}_j+
\hat{b}_i \hat{b}^{\dagger}_j) +\frac{1}{2}U\sum_i \hat{n}_i^{(b)}(\hat{n}^{(b)}_i-1)\nonumber\\
&&-\mu\sum_i \hat{n}^{(b)}_i\,,\label{eqn:bosehubbard}
\ea
where $t$ is the hopping strength, $U$ is the interaction strength, $\mu$ is the chemical potential, $\hat{b}_i$, $\hat{b}_i^{\dagger}$ are bosonic destruction/creation operators satisfying bosonic commutation relations on-site and commuting between sites, $\hat{n}_i^{(b)}$ is the bosonic number operator, and $\langle i,j\rangle$ denotes a sum over nearest neighbors.  The corresponding Fermi-Hubbard Hamiltonian in particle-hole symmetric form is
\ba
\hat{H_f}&=& -t\sum_{\langle i,j\rangle,\sigma}(\hat{f}^{\dagger}_{i\sigma} \hat{f}_{j\sigma}+
\hat{f}^{\dagger}_{j\sigma}\hat{f}_{i\sigma} )\nonumber\\
&&+U\sum_i \left(\hat{n}_{i\downarrow}^{(f)}-\frac{1}{2}\right)
\left(\hat{n}^{(f)}_{i\uparrow}-\frac{1}{2}\right)\nonumber\\
&&-\mu\sum_i \hat{n}^{(f)}_i\,,\label{eqn:fermihubbard}
\ea
where $\hat{f}_{i\sigma},\hat{f}_{i\sigma}^{\dagger}$ are fermionic destruction and creation operators satisfying fermionic anticommutation relations; $t$ and $U$ have the same meaning as in the bosonic case; and we have assumed $s$-wave interactions and spin 1/2, so that $\sigma\in\left\{\uparrow,\downarrow\right\}$.  We note that there is an exact mapping under the canonical transformation from the negative-$U$ to positive-$U$ Fermi-Hubbard Hamiltonian at half filling ($\mu=0$).  This will appear visually in our phase diagrams in Sec.~\ref{ssec:fermi}.

To find the ground states of these models, we use TEBD~\cite{tebdOpenSource} in 1D and propagate in imaginary time, starting with a state with uniform weights.  TEBD has one main convergence parameter, the number of eigenvalues $\chi$ retained in the reduced density matrix.  In all cases we have checked convergence by at least doubling $\chi$, and seeing if we obtain the same results in the plots; all plots have at least a $\chi$ of 16, and we have spot-checked a $\chi$ of 32.  We emphasize that as we have a very high resolution of ($250\times 100$) pixels in our diagrams, we cannot go to the preferred typical values of $\chi$ in the hundreds, in common use for TEBD calculations; also, grand canonical calculations are inherently more restrictive, due to the much larger state space.  We use both canonical and grand canonical ensembles depending on circumstances.  The latter is useful to understand the proper interpolation method to be used for the canonical ensemble for smaller numbers of sites.  In the canonical ensemble, chemical potential is derived by finite differencing, $\mu \approx E(N+1) - E(N)$; for just a few sites, this would lead to only a few points of resolution on the chemical potential axis without interpolation.  That is, a finite difference approximation for the chemical potential is only \emph{directly} useful for large $N$; interpolation can only be justified by comparing to the grand canonical results.  We illustrate this point in Sec.~\ref{ssec:ensemble}.
For fermions, there is an additional advantage of grand canonical calculations in that, in practice, we find that imaginary time propagation is less likely to get stuck in a local minimum.

For fermions, the local, or on-site dimension of the Hilbert space is fixed at $d=4$ since we consider only a single band.  For bosons $d$ provides a second convergence parameter.  In practice we truncate at $d=5+1$, meaning 0 to 5 bosons per site, and we have checked $d=7+1$ for specific points in the phase diagrams and found no visible changes.   We implement imaginary time propagation in the second order Suzuki-Trotter scheme to obtain the ground state.  Finally, we parallelize over the parameter space $(t/U, \mu/U)$ via MPI for bosons, and $(U/t, \mu/t)$ for fermions.  We typically perform our calculations on between $5\times 8$ and $60\times 8$ cores.  We note that TEBD becomes exact, aside from the $\mathcal{O}\left(\delta t^2\right)$ error of the Suzuki-Trotter decomposition, for $\chi \geq d^{\lfloor L/2 \rfloor}$.

Our quantum measures can be divided into four classes: moments, correlation, entropies, and fidelity.  Although these classes of measures can have significant overlap in the information they provide, nevertheless there are always cases where they are distinct.  For example, the spin singlet is maximally entangled but is not correlated.  Thus it is useful to consider all classes.

First, for moments, the energy, $E_b$, and filling factor, $\nu_b$, are defined as
\ba
E_b&=&\langle \hat{H}_b\rangle\,,\\
\nu_b&=&\frac{1}{L}\sum_{i=1}^L\langle \hat{n}^{(b)}_i\rangle\,,
\ea
where $L$ is the number of lattice sites.  To calculate $E_f$ and $\nu_f$ for fermions, one must also sum over the spin index $\sigma$.
%Also for fermions, the average local magnetization is given by
%\be M= \frac{1}{L}\sum_{i=1}^{L}\left\langle \left(\hat{n}^{(f)}_{i\uparrow} - %\hat{n}^{(f)}_{i\downarrow} \right)\right\rangle\,.
%\ee

Second, for correlations in bosons we calculated the quantum depletion, $D$, which is a measure of how far away the system is from the Penrose-Onsager definition of superfluidity~\cite{penrose1956}:
\be
D \equiv 1-\frac{\lambda_1}{\sum_{i=1}^{L}\lambda_i}\,,
\ee
where $\{\lambda_i\}$ are the eigenvalues of the single particle density matrix $\langle \hat{b}_i^{\dagger}\hat{b}_j\rangle$ ordered so that $\lambda_1 > \lambda_2 > \cdots > \lambda_L$.  For correlations in fermions we use three standard measures~\cite{scalettar1989,varney2009}.  The s-wave pairing correlation, $P$, is given by
\ba
P &=& \langle \hat{\Delta}^{\dagger}\hat{\Delta} + \hat{\Delta}\hat{\Delta}^{\dagger} \rangle\,,\\
\hat{\Delta}^{\dagger} &\equiv& \frac{1}{\sqrt{L}}\sum_{i=1}^{L}\hat{f}^{\dagger}_{i\uparrow}\hat{f}^{\dagger}_{i\downarrow}\,.
\label{eqn:pc}
\ea
The charge-density wave correlation, $C$, is given by
\ba
C &=& \langle \hat{\theta}\hat{\theta}^{\dagger}\rangle\,,\\
\hat{\theta}^{\dagger} &\equiv& \frac{1}{\sqrt{L}}\sum_{j=1}^{L}\sum_{\sigma} e^{i\pi j}\hat{n}_{j\sigma}^{(f)}\,.
\label{eqn:cdc}
\ea
The antiferromagnetic structure factor is given by
\ba
A&=&\frac{1}{L}\sum_{i,j=1}^L\left(-1\right)^{i-j}\langle \hat{S}^z_{i}\hat{S}^z_{i}\rangle\, ,\\
\hat{S}^z_{i}&\equiv& \frac{1}{2}\left(\hat{n}_{i\uparrow}^{(f)} - \hat{n}_{i\downarrow}^{(f)}\right)\,.
\ea

Third, for entropies, the average local von Neumann entropy~\cite{kitaev2006}, or average local entropy of entanglement, $\bar{S}$, is
\be
\bar{S}\equiv -\frac{1}{L}\sum_{i=1}^L\mathrm{Tr}\,\left(\hat{\rho}_i\log_d\hat{\rho}_i\right)\in[0,1]
\ee
where the local reduced density matrix, $\hat{\rho}_i$, is defined as
\be
\hat{\rho}_i = \mathrm{Tr}_{j\neq i}\hat{\rho}
\ee
and $\hat{\rho}$ is a pure-state density matrix for the whole system, as we work at zero temperature.  Note that the base $d$ of the log normalizes the entropy to lie between zero and unity.  We also studied the Q-measure~\cite{meyer2002,brennenGK2003} or average local impurity, $Q$, a special case of generalized entanglement~\cite{barnumH2003,barnumH2004}, which gives similar results to the von Neumann entropy (see Sec.~\ref{sec:bosefermi}):
\begin{equation}
Q\equiv \frac{d}{d-1} \left [1 - \frac{1}{L}\sum_{i=1}^L \mathrm{Tr}\,(\hat{\rho}_i^2) \right
] \in [0, 1]\,. \label{eqn:Qmeasure}
\end{equation}

Fourth, we calculated both fidelity, $f$, and fidelity susceptibility~\cite{buonsante2007,venuti2008,rigol2009b}, $\chi_f$.  The fidelity is an overlap measure on the states, and is given by
\be
f\left(\zeta,\zeta+\delta\zeta\right)\equiv\left|\langle
\psi\left(\zeta\right)|\psi\left(\zeta+\delta\zeta\right)\rangle\right|\,,
\label{eqn:fidelity}
\ee
where $\zeta$ is a control parameter in a quantum phase diagram, in our case either chemical potential or hopping.  The fidelity susceptibility is a second derivative of the fidelity in the control parameter, and we normalize it to the unit length:
\be
\chi_f\left(\zeta\right)\equiv\frac{1}{L}\frac{2\left(1-f\right)}{\left(\delta
\zeta\right)^2}\,,
\label{eqn:fs}
\ee
This proves to be a more useful quantity than the fidelity, as it is independent of the control parameter step size $\delta\zeta$, and a divergence of $\chi_f$ in the thermodynamic limit signals a quantum phase transition.
%For multiple parameters the fidelity susceptibility defines a Riemannian metric tensor over the parameter manifold:
%\be
%\left[\chi_f\right]_{\alpha \beta}=\Re\left(\langle \partial_{\alpha} \psi %|\partial_{\beta} \psi\rangle-\langle \partial_{\alpha} \psi|\psi\rangle\langle %\psi |\partial_{\beta} \psi\rangle\right)
%\label{eqn:hessian}
%\ee
%where $\Re$ denotes the real part and $\alpha,\beta$ are indices labeling the coordinates of the parameter manifold~\cite{venuti2007}.
Note that $\chi_f$ bears no relation to the convergence parameter for TEBD, $\chi$.  In the figures below we take $\zeta$ to be $t/U$ for the Bose-Hubbard Hamiltonian and $U/t$ for the Fermi-Hubbard Hamiltonian.

%In the grand canonical ensemble, one can consider a Hessian balancing changes in $\mu$ and $t$ as $\left[\chi_f\right]_{\mu\mu}\left[\chi_f\right]_{tt}-\left[\chi_f\right]_{\mu t}\left[\chi_f\right]_{t\mu}$, for example.  However, this leads to a trivial result of zero for the fidelity susceptibility within numerical precision.  In practice, it is preferable to calculate fidelity and fidelity susceptibility in the canonical ensemble so that changes in the conserved quantity, i.e., total number, do not dominate the picture.  Therefore $\zeta = t/U$ for fixed $N$ and $L$ in our studies, and there is no need to consider Eq.~(\ref{eqn:hessian}).

\section{Quantum Measures on Ground States of Hubbard Hamiltonians}
\label{sec:bosefermi}

\subsection{Bose-Hubbard Hamiltonian}
\label{ssec:bose}

%%%%%%%%%%% figure 2 %%%%%%%%%%%
%
\begin{figure*}
\begin{center}
\epsfxsize=\textwidth \epsfysize=\textwidth \leavevmode \epsfbox{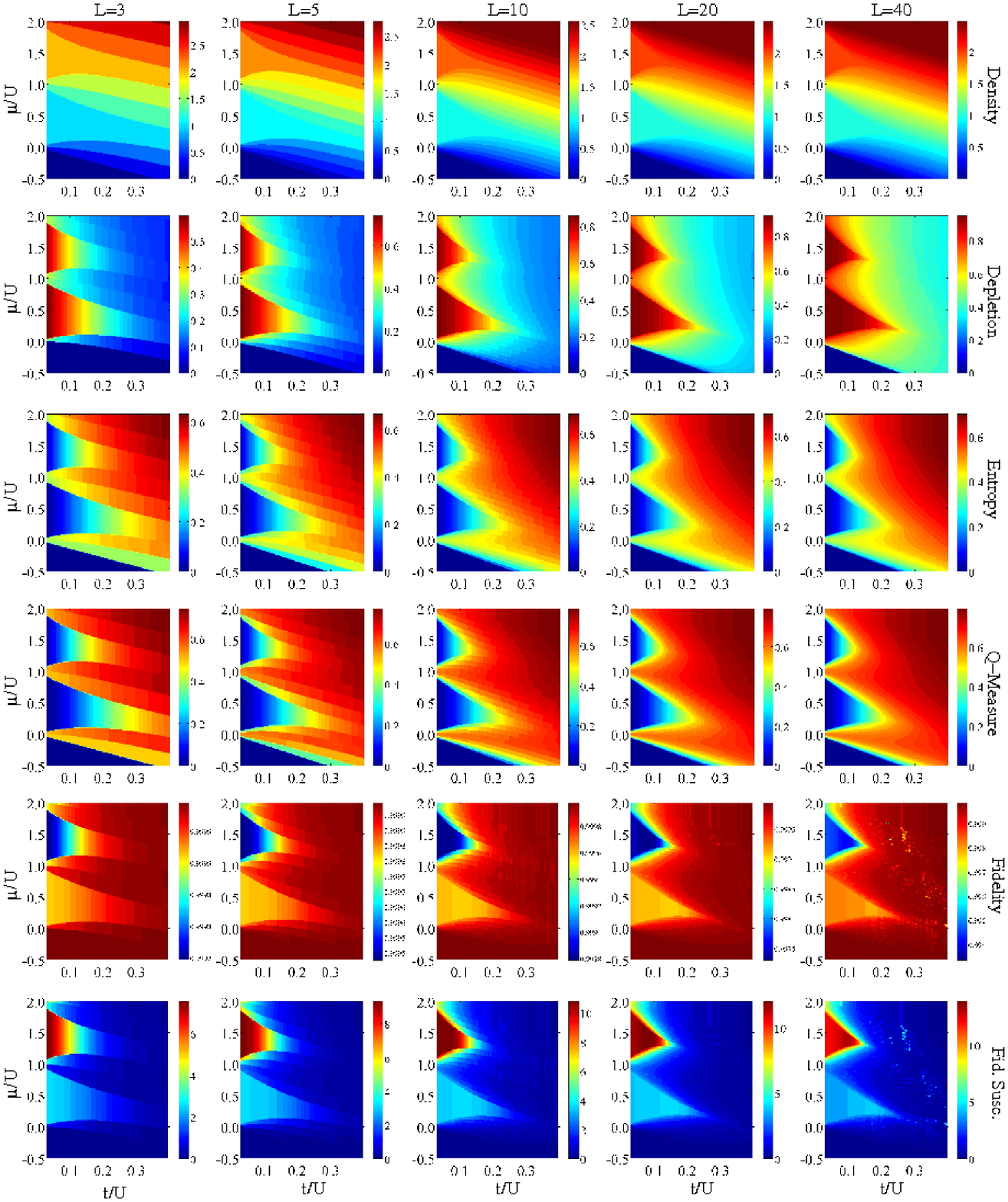}
\caption{\label{fig:boseAll} (color online) \emph{Quantum Measures for the Bose-Hubbard Hamiltonian}.  Shown is the density (first row), average local entropy of entanglement (second row), generalized entanglement or Q-measure (third row), quantum depletion (fourth row), fidelity (fifth row), and fidelity susceptibility (sixth row).  The panels show increasing numbers of sites from left to right: 3, 5, 10, 20, and 40.}
\end{center}
\end{figure*}

The Bose-Hubbard Hamiltonian phase diagram has been calculated in the mean field approximation in higher dimensions~\cite{fisher1989} and with both the DMRG method~\cite{monien1998,kuhner2000} and Monte Carlo in one dimension.  Trapped Bose systems with an external harmonic~\cite{rigol2009} and linear potential~\cite{carr2009j} have been treated with Monte Carlo and TEBD, respectively.  However, these studies did not focus on finite size effects without aiming at the thermodynamic limit.

In Fig.~\ref{fig:boseAll} are illustrated a suite of quantum measures for $L=3,5,10,20,$ and 40 sites, from left to right in column form.  The rows represent the different quantum measures as labeled on the right hand side: from top to bottom these are density, entropy, Q-measure, depletion, fidelity, and fidelity susceptibility.  The calculation of these measures and their definitions are given in Sec.~\ref{sec:equations}.  We emphasize that quantum Monte Carlo cannot be used to obtain entanglement.  Only matrix-product-state-based techniques can do so for general local Hamiltonians, including TEBD and DMRG~\cite{schollwock2005}, which lead to an equivalent result, although with different numerical implementations. Bethe-ansatz-based techniques have been used to calculate entanglement in certain limiting cases (near $U=0$) in the Fermi-Hubbard Hamiltonian~\cite{gu2004}, but not generally.  The case of $L=3$ is exact diagonalization for $\chi=15$ and $d=6$, but at $L=5$ the result is already approximate, since $6^2=36$ states would be required for exact diagonalization in matrix product state form (see Sec.~\ref{sec:equations}).  Each panel of Fig.~\ref{fig:boseAll} is of resolution $80 \times 240$ in $t/U \times \mu/U$.  The largest system of $L=40$ took several weeks to complete in the grand canonical ensemble on $20$ nodes with separate memory, each node consisting of 8 cores with shared memory, all on the Golden Energy Computing Organization~\cite{geco} high performance computing cluster ``Ra''.  For this reason we do not go to $\chi=32$ for the whole phase diagram, but only make spot checks, since simulations scale as $\chi^3$.  We calculated both canonical and grand canonical results, and show the canonical ones in Fig.~\ref{fig:boseAll}.

Turning to the physical interpretation of Fig.~\ref{fig:boseAll}, we begin with number.  In the canonical ensemble a definite total number is guaranteed.  This can be seen first in the average filling, which proceeds in multiples of $1/L$, as illustrated in the top row of the figure, starting from the vacuum and a filling of zero.  The vacuum region is evident in the lower portion of the panels in the top row of where it is shaded a deep blue.  The density increases in discrete jumps as the chemical potential is increased; interpolation is used between curves of definite number difference in order to fill in the picture, as described in Sec.~\ref{sec:equations} and Sec.~\ref{ssec:ensemble}.

In the second row we show the quantum depletion.  One clearly sees the highly depleted Mott lobes and the lowly depleted outer superfluid region.  Note that there are only two Mott lobes displayed in these figures; the apparent third, lowest such lobe is in fact the vacuum, $N=0$, for which quantum depletion is ill-defined.
In keeping with the 1D nature of the problem we consider, we observe that even the superfluid region is still depleted at a level of about 30\,\% far away from the Mott lobes.  Again, the smooth nature of the Mott-insulator-superfluid transition near the tip is in strong contrast to the sharper transition in other regions.

In the third and fourth rows of Fig.~\ref{fig:boseAll} we show entropies.  For completeness we compare the von Neumann entropy or entropy of entanglement, which is now a standard measure used in quantum information~\cite{nielsenMA2000} and also proposed for the study of quantum phase transitions~\cite{osterlohA2002,kitaev2006,dengS2008}, to the Q-measure, which has been more closely connected to local observables~\cite{barnumH2003,sommaR2004}.  We find both measures useful and nearly but not exactly the same; we choose the Q-measure for the rest of our study.  Note that the local von Neumann entropy of the Mott insulating regions is low by this measure, which reflects the nearly pure number state on each site in the Mott phase.  On the other hand, the superfluid region displays high entropy, because the superfluid state is delocalized across sites, leading to mixed states on each site when measured separately.  This contrasts with the usual association of low thermal entropy with a superfluid phase.

In the second through fourth rows one can clearly pick out the Mott-insulating region.  It has a sharp phase boundary for small $t/U\lesssim 0.2$.  K\"uhner, White and Monien~\cite{kuhner2000} first provided an accurate evaluation of the critical value, $t_c=0.29 \pm 0.1$, for the Kosterlitz-Thouless (KT) transition at the tip of the Mott lobe in the infinite system limit; in our notation this corresponds to a value of $t/U=0.29$.  Near $t_c$ the entropy changes more continuously for 3, 5, and 10 sites, while for 20 sites the lobe begins to close.  Finally at 40 sites, on the far right of the figure, the region of continuous change is very narrow.  We clarify that by ``sharper boundary'' as compared to ``more continuous'' we are speaking in relative terms, specifically of the difference in entropies between states with commensurate and incommensurate filling.  There is no truly sharp transition since this is a finite system; all ``quantum phase transitions'' seen here are in fact formally crossovers between regions with different characteristics.

Finally, in the fifth and sixth rows we show the fidelity $f$ and fidelity susceptibility $\chi_f$.  In our present implementation we compare the neighboring pixels of the figures in $t/U$ for curves of fixed $N$ and $L$, as described in Sec.~\ref{sec:equations}.  This excludes a comparison between changes in total number; such a comparison, which we also performed in the grand canonical ensemble, is not shown, because features in $f$ and $\chi_f$ are dominated by changes in the conserved quantity.  In the thermodynamic limit the superfluid region of the phase diagram forms a continuous tensor product space and the fidelity is everywhere zero even for arbitrarily small $\delta\zeta$.  This is the well-known \emph{Anderson orthogonality catastrophe}~\cite{anderson1967}.  Thus, the fidelity, through its scaling and qualitative behavior, reflects for a finite system the critical behavior of an infinite system.  Indeed, the Mott lobes appear clearly in this set of measures, as can be seen in the figure.

\begin{figure}
\begin{center}
\epsfxsize=8cm \epsfysize=13 cm \leavevmode \epsfbox{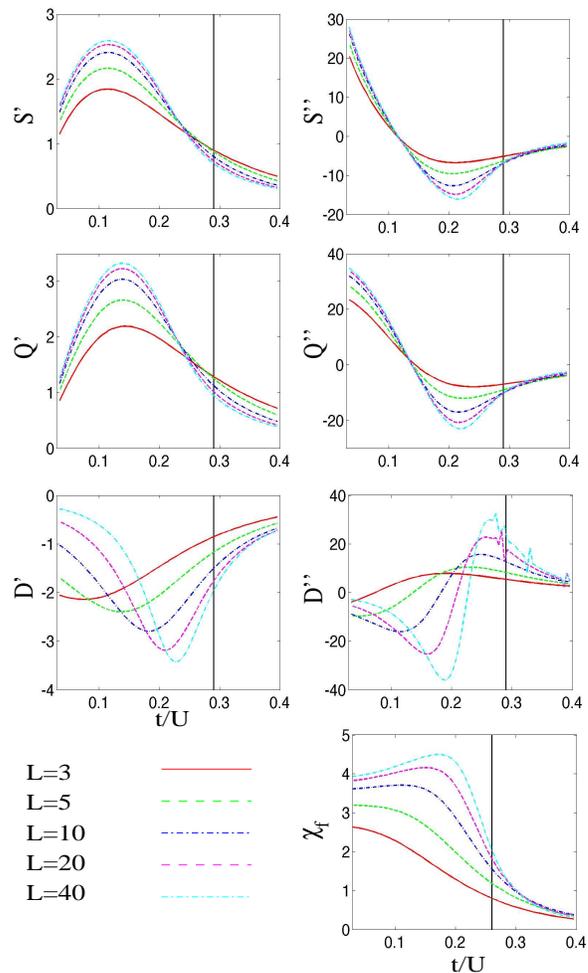}
\caption{\label{fig:estimators} (color online) \emph{Crossover estimators}.  First derivative (left column) and second derivative (right column) with respect to $t/U$ of various quantum measures for unit filling in the canonical ensemble.  The vertical line corresponds to the known value of $t_c$.}
\end{center}
\end{figure}

In order to provide a more \emph{quantitative} estimate of the tip location for finite systems than a by-eye estimation from Fig.~\ref{fig:boseAll}, we study quantum measures and derivatives thereof for unit filling $N=L$ in the canonical ensemble.  Three figures illustrate our results.  In Fig.~\ref{fig:estimators} are shown derivatives of quantum measures which could be used to define a crossover; in rows from top to bottom these are derivatives of average local von Neumann entropy, Q-measure, and quantum depletion; and lastly, the fidelity susceptibility.  The fidelity susceptibility provides a particularly sharp signature of the Mott phase, a result that was noted previously in the context of perodic boundary conditions and the completely connected graph~\cite{buonsante2007}.  Within the $n^{\mathrm{th}}$ Mott lobe, $\chi_f$ assumes values close to those given by second order perturbation theory~\cite{rigol2009b} in the limit that $t/U\to 0^+$:
\be
\chi_f = 2n(n+1)\frac{L-1}{L}
\label{eqn:perturb}
\ee
for open boundary conditions.  For periodic boundary conditions the system size cancels and the expression is $\chi_f = 2n(n+1)$.  These perturbations represent particle-hole fluctuations, which are weighted more heavily in higher Mott lobes due to Bose stimulation.  This causes $\chi_f$ to have a higher value in the upper Mott lobe in Fig.~\ref{fig:boseAll} as compared to the lower lobe.

\begin{figure}
\begin{center}
\epsfxsize=8cm \epsfysize=6.5 cm \leavevmode \epsfbox{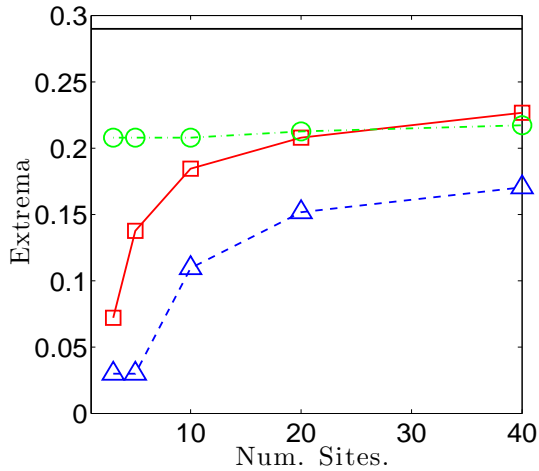}
\caption{\label{fig:tip} (color online) \emph{Estimates of the crossover}.  Extrema of the first derivative of the depletion (red squares), the fidelity susceptibility (blue triangles), and the second derivative of the Q-measure for unit filling in the canonical ensemble (green circles).  All derivatives are with respect to $t/U$.  The horizontal line again corresponds to the limit $t_c$.  The curves are a guide to the eye.}
\end{center}
\end{figure}

\begin{figure}
\begin{center}
\epsfxsize=8cm \epsfysize=6.5 cm \leavevmode \epsfbox{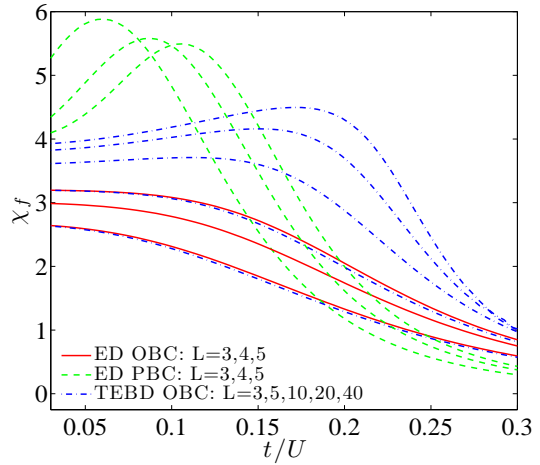}
\caption{\label{fig:exactDiagonalization} (color online) \emph{Open vs. Periodic Boundary Conditions}.  Fidelity susceptibility for fixed $N=L$ in a comparative study of open (OBC) and periodic boundary conditions (PBC); the latter can only be obtained via exact diagonalization (ED) in our present code, while OBCs can be compared directly with TEBD.  We note that (1) PBC have a substantially different fidelity for smaller systems, with a peak at much smaller $t/U$, and (2) TEBD and exact diagonalization match up to the $\mathcal{O}\left(\delta t^2\right)$ error from the Suzuki-Trotter decomposition for small systems, validating our code.}
\end{center}
\end{figure}

As the number of sites increases from 3 to 40
%(eventually go to 80 or 100 with follow-up calculations)
the extrema move towards the thermodynamic limit $t_c$.  Figure~\ref{fig:tip} summarizes the most useful extrema of these curves in an easy to read single plot showing that quantum depletion and fidelity susceptibility appear to converge most rapidly to their asymptotic values in the thermodynamic limit.  However, $\chi_f$ is affected by the boundary conditions, as can be seen in Eq.~\ref{eqn:perturb} as well as in Fig.~\ref{fig:exactDiagonalization}, where we compare periodic and open boundary conditions using both exact diagonalization and TEBD; this was also a significant issue in DMRG and perturbation theory studies, where earlier calculations~\cite{monien1998} obtained a less accurate estimate of the critical point due to the choice of boundary conditions~\cite{kuhner2000}.  This figure also serves as a check on our code.  We emphasize that open boundary conditions are closer to the conditions of ultracold atom experiments at the present time.

\subsection{Fermi-Hubbard Hamiltonian}
\label{ssec:fermi}

%%%%%%%%%%% figure 3 %%%%%%%%%%%
%
\begin{figure*}[t]
\begin{center}
\epsfxsize=\textwidth \epsfysize=\textwidth \leavevmode \epsfbox{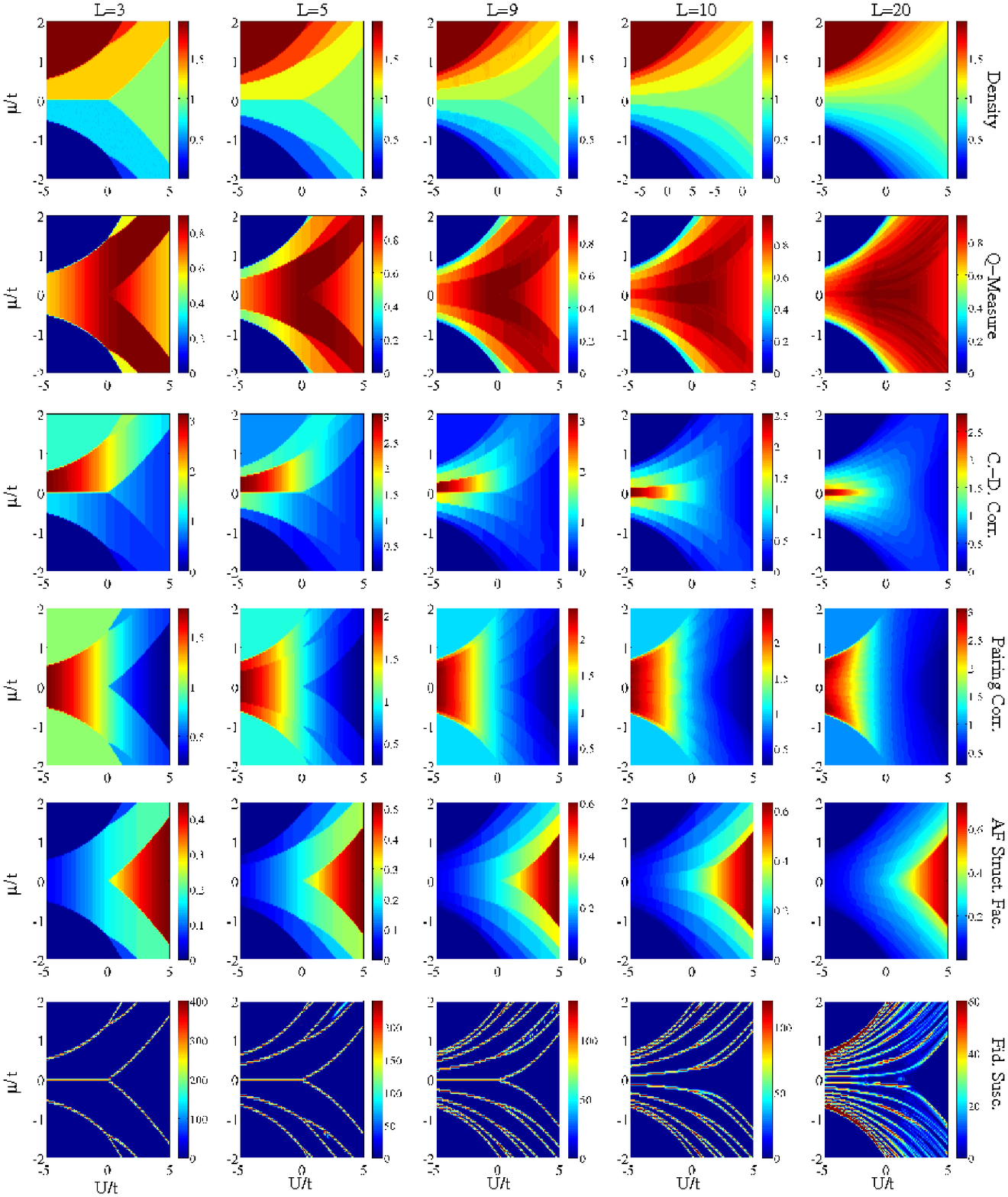}
\caption{\label{fig:fermiAll} (color online) \emph{Quantum Measures for the Fermi-Hubbard Hamiltonian in Particle-Hole Symmetric Form}.  Density (first row), generalized entanglement or Q-measure (second row), charge-density wave correlation (third row), pairing correlation (fourth row), antiferromagnetic structure factor (fifth row), and fidelity susceptibility (sixth row).  The panels show increasing numbers of sites from left to right: 3, 5, 9, 10, and 20.}
\end{center}
\end{figure*}

The Fermi-Hubbard Hamiltonian in 1D has an exact solution~\cite{Lieb1968,Lieb2003} via the Bethe ansatz; in principle, one can extract any information from such a solution, including all measures of interest, and the Bethe ansatz has been applied to finite as well as infinite domains.  Because of their computational complexity, Bethe ansatz results are not widely used for performing partial traces of the type that we consider here~\cite{gu2004}.  Therefore it is useful to consider finite size effects for a suite of measures simultaneously, including measures not previously accessed through exact solution techniques.  TEBD or DMRG methods also enable one to ``turn on'' non-mean-field effects, i.e. entanglement, in discrete steps in the form of the parameter $\chi$, as we show in Sec.~\ref{ssec:meanfield}.

Figure~\ref{fig:fermiAll} displays ground state solutions of the Fermi-Hubbard Hamiltonian  [Eq.~(\ref{eqn:fermihubbard})] in the grand canonical ensemble for 3, 5, 9, 10, and 20 sites from left to right in columns, following the same pattern as Fig.~\ref{fig:boseAll} but with measures appropriate to fermions.  Since we take $\chi=16$, for 3 and 5 sites TEBD is equivalent to exact diagonalization, as $d=4$; in fact, we explicitly checked our results with a separate exact diagonalization routine.  The first row shows the density.  For positive $U$ we observe that all number states are allowed; as in Sec.~\ref{ssec:bose}, the grand canonical ensemble settles on a total-number-conserving state for these small systems due to the discreteness of the spectrum.  For example, for $L=3$ there are seven colored regions for increasing $\mu/t$ and $U/t$ greater than 0, of which all regions are visible up to $U/t \approx 2$; these correspond to 0,1,$\ldots$,6 fermions.  For $L=5$ there are eleven colored regions, of which 2 are barely visible due to our choice of scale, and for higher $L$ we observe the same pattern.  For odd numbers of sites the exact mapping from positive $U$ to negative $U$ in Eq.~(\ref{eqn:fermihubbard}) at zero chemical potential does not hold, and therefore there is no symmetry around the vertical axis at $U/t=0$ for $L=3,5,9$, while for $L=10,20$ we observe this symmetry.  For negative $U$, particles enter only in pairs, excepting at $\mu=0$ for $L$ even, as can be seen from the figure.

In the second row of Fig.~\ref{fig:fermiAll} we show the entanglement entropy.  One clearly observes the particle-hole symmetry between negative and positive chemical potential.  For odd numbers of sites, the most highly entangled regions for positive $U$ are those in which there is one extra particle or hole above half filling; for even systems it is two extra particles or holes, as can be seen in the stripe-like pattern for $L=20$, for example.  States nearer the band insulator, with just a few particles (holes) compared to the vacuum (maximal filling) are less entangled.  Entanglement also decreases in the Mott region for increasing $U$ since the system moves closer to a product state.  There is a discontinuity in the $Q$-measure between regions of half filling and slightly away from half filling which vanishes at $U$=0.  This can be understood as the requirement that the derivatives approaching from above and below are required to satisfy $dQ/dn|_{n=1^+}=-dQ/dn|_{n=1^-}$ by particle-hole symmetry.  For negative $U$ the most highly entangled regions are near zero chemical potential and near half-filling.  Therefore, in general the pattern for both positive and negative $U$ is that the entanglement is largest near half filling.  We show only the Q-measure, having established in Sec.~\ref{ssec:bose} that the average local von Neumann entropy does not show substantial additional information.

In the third row of Fig.~\ref{fig:fermiAll} we show the charge-density correlation (CDC).  This measure shows how close the system is to a ``checkerboard'' occupying every other site.  For an odd number of sites, as visible in $L=3,5,9$, and negative $U$, the CDC is higher for two pairs than for one pair, by the nature of the measure [see Eq.~(\ref{eqn:cdc})] -- this is due to the fact that there is one extra pair which cannot be placed.  For even numbers of sites, as seen in $L=10,20$ it is particle-hole symmetric, as expected, yielding the same result upon flipping all plots around the $\mu=0$ horizontal axis.  CDCs do not follow the positive-$U$ to negative-$U$ mapping of the Fermi-Hubbard Hamiltonian.  CDCs decrease as $U>0$ is increased, since the system is closer to a Mott product state. In the fourth row of Fig.~\ref{fig:fermiAll} we show the pairing correlations, which are of interest only in the negative-$U$ region where pairing naturally occurs due to the negative interaction.

In the fifth row of Fig.~\ref{fig:fermiAll}, we show the antiferromagnetic structure factor.  This measure is the spatial fourier transform of the spin-spin correlation function evaluated at $q=\pi$, and is motivated by the fact that a canonical transformation maps the half-filled positive-$U$ Hubbard model in the strong coupling limit $U/t\gg 1$ onto a spin-1/2 antiferromagnetic Heisenberg model with $J=4 t^2/U$ in an external magnetic field $h=2\mu$
\ba
\hat{H}=J\sum_{\langle i,j\rangle}\mathbf{S}_i\cdot\mathbf{S}_j-h\sum_i \hat{S}^z_{i}\, .
\ea
This measure jumps sharply at half filling where the antiferromagnetic order is expected to exist, grows positively and reaches a maximum for some value of $U$, and then decreases to 0 for large $U$.  The symmetry about $\mu$ is preserved because the operators involved are $\hat{S}_z^{\left(i\right)}=\frac{1}{2}\left(\hat{n}_{i \uparrow}^{\left(f\right)}-\hat{n}_{i\downarrow}^{\left(f\right)}\right)$ which vanish for on-site singlet pairs.

In the sixth row of Fig.~\ref{fig:fermiAll} we show the fidelity susceptibility.  Due to our use of the grand canonical ensemble, only different total number states have an overlap significantly different from unity for small systems.  For larger systems, fidelity susceptibility becomes less dominated by differences in number states as the number of jumps in number approaches the resolution in $(\mu/t,U/t)$, as was also the case for the Bose-Hubbard Hamiltonian in Fig.~\ref{fig:boseAll}.
%this may change if we are able to make the canonical ensemble work with imaginary time propagation

\section{Key Numerical Comparisons}
\label{sec:compare}

\subsection{Grand Canonical vs. Canonical Ensembles}
\label{ssec:ensemble}

%%%%%%%%%%% figure 4 %%%%%%%%%%%
%
\begin{figure}
\begin{center}
\epsfxsize=8cm \epsfysize=6cm \leavevmode \epsfbox{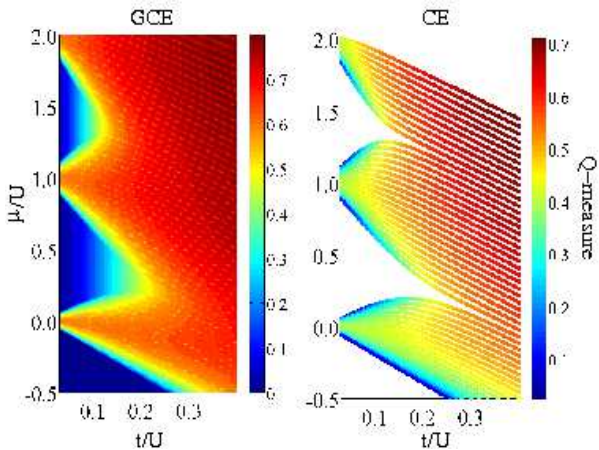}
\caption{\label{fig:ensemble} (color online) \emph{Comparison of Canonical Ensemble (CE) and Grand Canonical Ensemble (GCE).}  Shown is the Q-measure (generalized entanglement) for the Bose-Hubbard Hamiltonian and $L=20$ sites.  The CE is restricted to curves of fixed number difference.  To obtain the images in Fig.~\ref{fig:boseAll}, we interpolate the CE by taking the nearest fixed line and thus ``filling in'' the phase diagram.  The GCE shows that this method obtains the correct result, thus overcoming the finite differencing error inherent to the CE: $\mu\approx E(N+1)-E(N)$.  Small qualitative differences remain between the two ensembles, as can be observed by comparing to the corresponding panel (second row, fourth column) in Fig.~\ref{fig:boseAll}.}
\end{center}
\end{figure}

%%%%%%%%%%% figure 4 %%%%%%%%%%%
%
\begin{figure*}
\begin{center}
\epsfxsize=\textwidth \epsfysize=0.75\textwidth \leavevmode \epsfbox{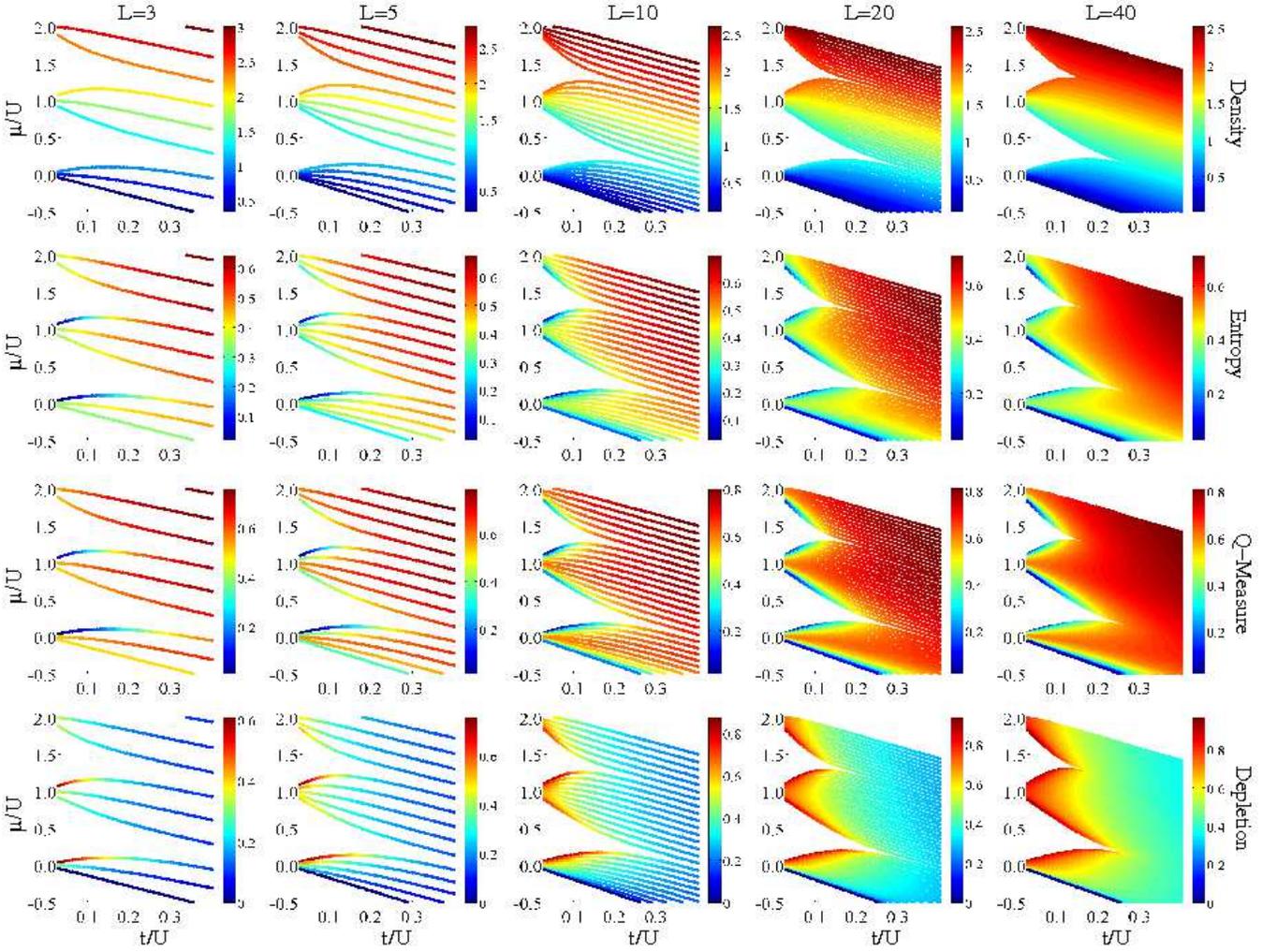}
\caption{\label{fig:fullCE} \emph{Quantum Measures for the Bose-Hubbard Hamltonian in the Canonical Ensemble.}  This figure shows the actual data underlying Fig.~\ref{fig:boseAll}; for smaller systems, the canonical ensemble provides very poor resolution in chemical potential, as can be observed in the first few columns.  As the system size increases, the resolution in chemical potential goes up and the Mott lobe tip appears.}
\end{center}
\end{figure*}

The canonical ensemble is much preferable for large systems because it is more efficient.  However, in order to obtain a chemical potential one must approximate the derivative in $\mu \equiv (\partial E / \partial N)|_V$.  Our volume is automatically fixed by performing the derivative for a given system size $L$.  The derivative is then evaluated by finite differences, either a simple forward difference of $\mu \approx E(N+1)-E(N)$, or a higher order finite difference.  At any level of finite difference approximation, for smaller systems there is only a very low resolution for the chemical potential axis.  This leads to a lack of information about quantum measures.  This issue is highlighted in Fig.~\ref{fig:ensemble} by showing what the actual data is for the canonical ensemble (right panel) as compared to the grand canonical ensemble (left panel) for 20 sites.  In the canonical calculation, each horizontal curve represents an allowed value of $N$.  The grand canonical ensemble allows us to motivate the interpolation of the canonical ensemble result to go past the finite-difference result and obtain a continuous plot.  Our interpolation algorithm is to take the value from the nearest curve according to the standard distance measure $\sqrt{(\mu_2-\mu_1)^2-(t_1-t_2)^2}$.  Figure~\ref{fig:fullCE} shows the actual canonical ensemble data used to produce Fig.~\ref{fig:boseAll}.

The initial state chosen for our imaginary time propagation has the same projection onto every Fock state in the Hilbert space allowed by our numerical truncation.  In our grand canonical calculations, this includes states which do not have the same total number, $N$.  However, for small systems, number states are sufficiently different in energy that imaginary time propagation converges on a state with a definite $N$.  On the border between number states one requires a very high $\chi$ to deal with the near degeneracy of states with different $N$.  This leads to the regular pattern of bright spots seen in the left panel of Fig.~\ref{fig:ensemble}.  The sole effect of higher $\chi$ is to remove bright spots from the entropy and Q-measure.  The density of such low-entropy spikes increases with increasing $L$ because a higher $\chi$ is needed for a larger system -- more sites means the possibility for more spatial entanglement.  This is another reason why canonical ensemble calculations are preferred for bosons.  We have verified that all panels in Fig.~\ref{fig:boseAll} match the grand canonical picture, with the exception of the bright spots, which do not occur in canonical ensemble calculations.

Although we do not illustrate it in Fig.~\ref{fig:ensemble}, the variance in the total number, $\langle \hat{N}^2\rangle - \langle \hat{N}\rangle^2$, where $\hat{N} \equiv \sum_i \hat{n}^{(b)}_i$, is zero to 5 or 6 digits of precision.  This is in fact a measure of convergence for our grand canonical simulations; we require all Schmidt values (eigenvalues of the reduced density matrix) be converged to 10 digits in imaginary time propagation, and obtain this total number variance, having started with an initial state with weights on all total numbers allowed by the cut-off $d$, i.e., from 0 atoms to $\left(d-1\right)\times L$ atoms.

\subsection{Gutzwiller Mean Field vs. Entangled Time-Evolving Block Decimation}
\label{ssec:meanfield}

%%%%%%%%%%% figure 5 %%%%%%%%%%%
%
\begin{figure*}
\begin{center}
\epsfxsize=0.8\textwidth \epsfysize=\textwidth \leavevmode \epsfbox{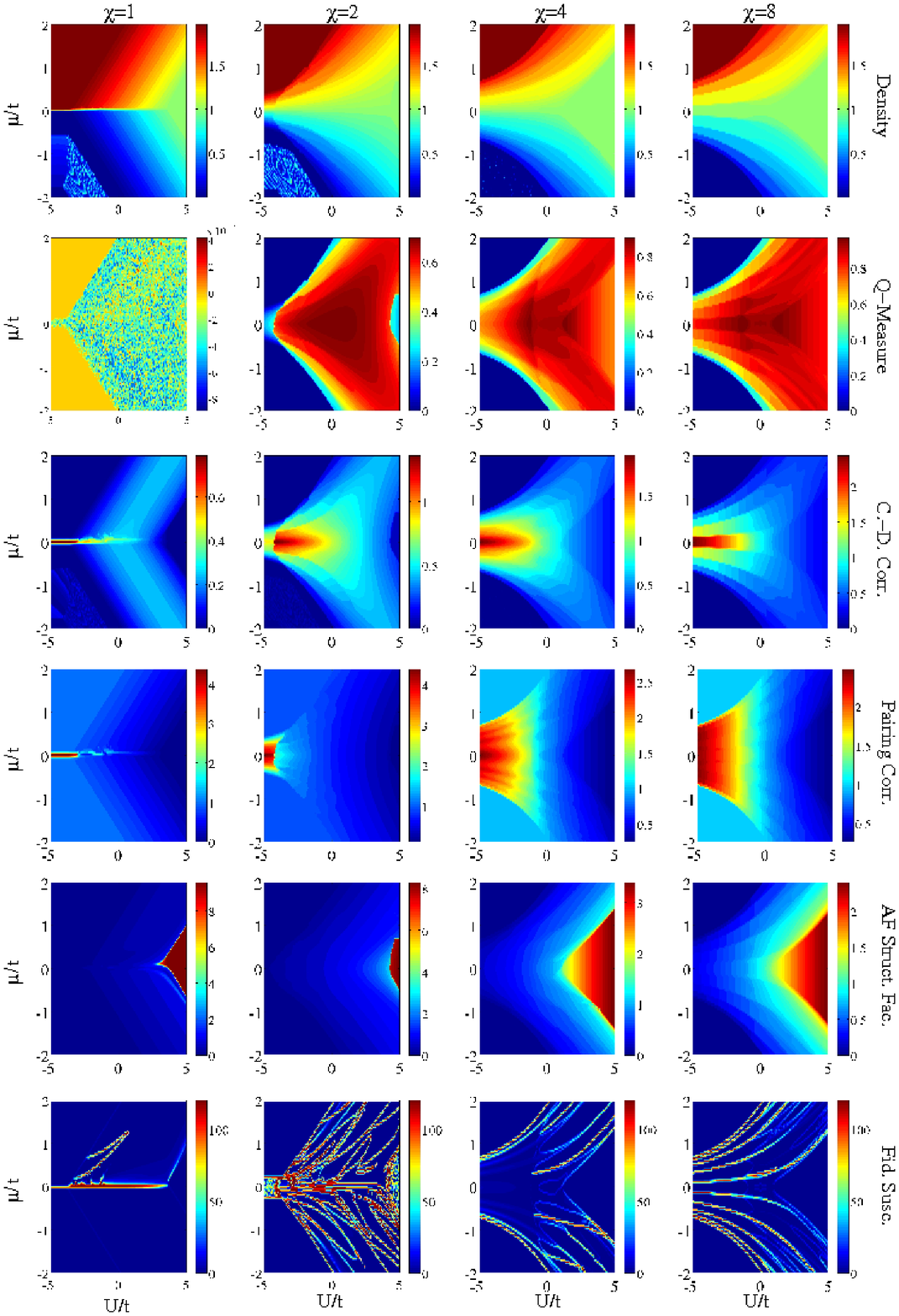}
\caption{\label{fig:chi} (color online) \emph{Increasing Corrections to Mean Field Gutzwiller Ansatz}.  The same quantities as calculated in Fig.~\ref{fig:fermiAll} for the Fermi-Hubbard Hamiltonian, but for 10 sites and in increasing values of $\chi$ from left to right: 1,2,4,8.  $\chi=1$ corresponds to the Gutzwiller mean field approximation.  The speckle structure observed in the vacuum region and for entanglement in the $\chi=1$ case is simply numerical noise associated with the calculation of small numbers.}
\end{center}
\end{figure*}

Since mean field methods are usually the most tractable approaches to solution of many body problems, identifying their domain of validity is important.  For instance, for the three-dimensional Bose-Hubbard Hamiltonian one can obtain the phase diagram via the Hubbard-Stratonovich transformation~\cite{fisher1989}, which is a mean field method, or alternately, via the Gutzwiller mean field approximation, as discussed in Sec.~\ref{sec:introduction}.  On the other hand, the Gutzwiller method fails for the one-dimensional Bose-Hubbard Hamiltonian~\cite{lewensteinM2007}.  It is natural to ask, how far beyond mean field does one need to go in order to obtain correct results for essential quantum measures?

To address this question, we repeat the calculations of Sec.~\ref{ssec:fermi} for the Fermi-Hubbard Hamiltonian for 10 sites only, starting with the mean field calculation of $\chi=1$ and then doubling $\chi$ successively.  Figure~\ref{fig:chi} shows the results: each column corresponds to $\chi=1,2,4,8$ while the rows correspond to the same quantum measures as Fig.~\ref{fig:fermiAll}.  Note that $\chi=16$ can be found in the third column of Fig.~\ref{fig:fermiAll}, so we do not repeat it here.  We make several remarks concerning this figure.  The density is a first moment and only requires lower values of $\chi$, as can be seen in the first row.  For $\chi=1$ in the first column, the entanglement shown in the second row is zero to 16 digits of accuracy, reflecting our use of double precision floating point arithmetic in our implementation of TEBD.  The entanglement does not converge until $\chi=16$, as can be determined from a comparison of Fig.~\ref{fig:chi} and Fig.~\ref{fig:fermiAll}.

In the third row, for large positive $U/t$ we obtain the correct result for charge density correlations at half-filling even for very low $\chi$ because the Mott phase is nearly a product state.  The pairing phase for negative $U/t$, on the other hand, requires higher $\chi$, as can be seen in the fourth row.
This can be understood from the fact that superfluids and superconductors are highly entangled in position space; similarly, in the Bose Hubbard Hamiltonian one must go to high $\chi$ in order to resolve the superfluid region far away from the Mott lobes.  By ``spatial entanglement'' we refer to exactly the Schmidt number used in TEBD.  The fifth row shows anti-ferromagnetic order.
%add description here of anti-ferromagnetic chi dependence -- one line

Finally, in the sixth row we observe that the main features of the fidelity susceptibility appear already for $\chi=2$, because fidelity measures in the grand canonical ensemble are mainly sensitive to total number states in our system; correct total number states are obtained at very low $\chi$.  We observe that particle-hole symmetry is maintained even at the mean field level of $\chi=1$, while the positive-$U$ to negative-$U$ mapping at half-filling and zero chemical potential is only good for at least $\chi=2$.

\begin{figure}
\begin{center}
\epsfxsize=8cm \epsfysize=8cm \epsfbox{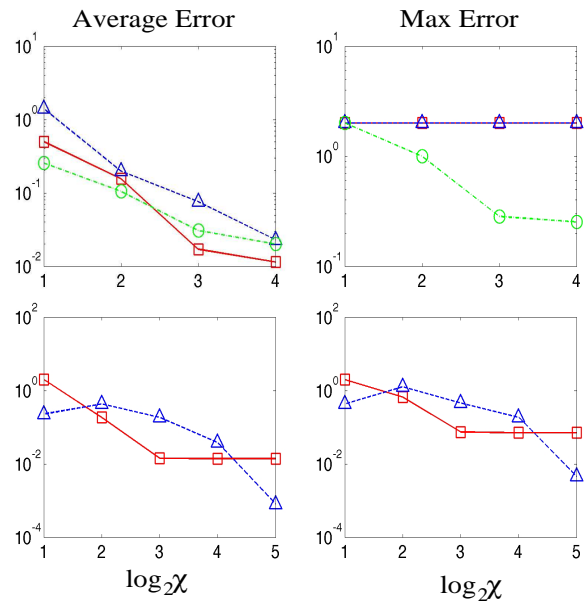}
\caption{\label{fig:error} (color online) \emph{Quantitative Error Analysis for increasing entanglement ($\chi$) in TEBD}.  Average error (left column) and maximum error (right column) for the Fermi-Hubbard Hamiltonian (top row) and Bose-Hubbard Hamiltonian (bottom row) as a function of increasing powers of $\chi$.  For fermions, we show the quantum measures of density (red squares), Q-measure or generalized entanglement (blue triangles), and pairing correlations (green circles); for bosons, we show the error in the Q-measure (red squares) and quantum depletion (blue triangles) in the grand canonical ensemble. Curves are a guide to the eye only.}
\end{center}
\end{figure}

Although Fig.~\ref{fig:chi} shows that $\chi=16$ is sufficient to obtain by-eye converged values of all quantum measures considered in Fig.~\ref{fig:fermiAll}, a more careful quantitative study is desirable.  In Fig.~\ref{fig:error} we show this study for both fermions and bosons for 10 sites.  Both the average error (left panels) and maximum error (right panels) over all pixels in $(\mu/t,U/t)$ for fermions and $(\mu/t,t/U)$ for bosons are evaluated.  The error for any given pixel $p$ is defined as
\be
\varepsilon_p(M) \equiv \left|\frac{M_{p}(2^j)-M_{p}(2^{j-1})}{\frac{1}{2}[M_{p}(2^j)+M_{p}(2^{j-1})]}\right|\,,
\ee
where $M(2^j)$ is the quantum measure of interest evaluated at Schmidt number $\chi=2^j$ and $j$ is an integer, in our case taken to be 1, 2, 3, 4, 5.  That is, we evaluate measure $M$ at $\chi$ of 16, 8, 4, 2 and compare it to the evaluation of the same measure at the previous smaller value of $\chi$.  The maximum and average errors are given by
\ba
\varepsilon_{\mathrm{max}}(M)&\equiv&\max_p \varepsilon_p(M)\,,\\
\bar{\varepsilon}(M)&\equiv&\frac{\sum_p \varepsilon_p(M)}{\sum_p 1}\,.
\ea
For fermions we calculate these two errors for density, Q-measure, and pairing in the grand canonical ensemble for 10 sites, corresponding to Fig.~\ref{fig:chi}.  For brevity, we do not show an analogous figure to Fig.~\ref{fig:chi} for bosons, but calculate maximum and average error for $L=10$ sites and the grand canonical ensemble shown in Fig.~\ref{fig:ensemble} in the third column.  As can be seen from the figures, the average error is about 1\,\% for $\chi=16$ for bosons, and the maximum error is about 10\,\%.  For fermions it somewhat larger for the same $\chi$.  The maximum error appears large, but upon a detailed investigation we find that this error is localized in the vacuum or band insulator regions; this means that the max error is dominated by imaginary time propagation effects.  Other regions have an error corresponding to the average error, including the important half-filling Mott region for positive $U$.  In general, local measures like density or entanglement converge faster than non-local measures like quantum depletion, as can be seen in the figure.  Also, the bosonic measures converge more quickly than do the fermionic measures.

%%%%%%%%%%% figure 6 %%%%%%%%%%%
%

\section{Conclusions}
\label{sec:conclusion}

We explored finite size effects in the Bose- and Fermi-Hubbard Hamiltonians with TEBD and imaginary time propagation, studying a range of quantum measures including moments, correlations, entanglement, and fidelity.  We found that entanglement~\cite{nielsenMA2000,weiTC2005} followed the same pattern as more traditional measures for quantum phase transitions, such as depletion for the Bose-Hubbard Hamiltonian and the superfluid-Mott-insulator transition, and pairing, charge-density, and antiferromagnetic structure factor for the Fermi-Hubbard Hamiltonian in the Mott transition; the exception was the tip of the Mott lobe, which was best described by the depletion, in terms of converging to the thermodynamic result more quickly as a function of system size.
Fidelity susceptibility provides a useful alternative measure for characterizing the phase transition, and is a non-local measure.

We compared results computed within the grand canonical ensemble and canonical ensembles: the chemical potential can be approximated in the canonical ensemble by a finite difference, a method only useful for large $N$.  However, by first performing a grand canonical calculation and then using it to motivate an interpolation of the canonical ensemble we could subsequently use the more efficient canonical calculations to obtain more highly converged results.  The canonical ensemble is also more useful for fidelity susceptibility calculations, which are otherwise dominated by changes in the conserved quantity of total number.

We performed quantitative convergence studies and showed that particle-hole symmetry is maintained in the particle-hole symmetric form of the Fermi-Hubbard Hamiltonian for all values of $\chi$, but the positive-$U$ to negative-$U$ mapping at half-filling and zero chemical potential is correct only for at least $\chi=2$, meaning it is not satisfied in the mean field approximation.  We found that for bosons the tip of the Mott lobe is wide-open for smaller system sizes of less than about 20 sites, so that exploring quantum phase transitions by changing chemical potential finds sharp effects while changing hopping find a very broad crossover.

We acknowledge useful discussions with Ippei Danshita, Ludwig Mathey, Ryan Mishmash, Alejandro Muramatsu, and Marcos Rigol.  This work was supported by the National Science Foundation under Grant PHY-0547845 as part of the NSF CAREER program (LDC, MLW) and by the Aspen Center for Physics (LDC), by the Summer Undergraduate Research Fellow (SURF) program at the National Institute of Standards and Technology (DGS, RCB), and by the NSF under Physics Frontiers Center grant PHY-0822671 (RCB, CWC).

%\bibliographystyle{prsty}
%\bibliography{refs,../../../refs/refs}

\end{document}